\documentclass[aps,preprint,superscriptaddress,floatfix,amsmath,amssymb]{revtex4}
\usepackage{inputenc}
\usepackage[a-1b]{pdfx}
\usepackage{amsmath}
\usepackage{amssymb}
\usepackage{amsthm}
\usepackage{bm}
\usepackage{graphicx}
\usepackage{subfig}
\usepackage{hyperref}
\usepackage[capitalise]{cleveref}

\begin{document}

%\preprint{AIP/123-QED}

\title{Community Formation in Wealth-Mediated Thermodynamic Strategy Evolution}
% Force line breaks with \\
\author{Connor Olson}
\email{cjo5325@psu.edu}
\affiliation{Dept. of Mathematics, Penn State University, University Park PA 16802}

\author{Andrew Belmonte}%
\email{andrew.belmonte@gmail.com}
\affiliation{Dept. of Mathematics and Huck Institute, 
Penn State University, University Park PA 16802}

\author{Christopher Griffin}
\email{griffinch@psu.edu}
\affiliation{%
Applied Research Laboratory, Penn State University, University Park PA 16802
}%

\date{June 26, 2022}% It is always \today, today,
             %  but any date may be explicitly specified

\begin{abstract}
We study a dynamical system defined by a repeated game on a 1D lattice, in which the players keep track of their gross payoffs over time in a bank. Strategy updates are governed by a Boltzmann distribution which depends on the neighborhood bank values associated with each strategy, relative to a temperature scale which defines the random fluctuations. Players with higher bank values are thus less likely to change strategy than players with lower bank value. For a parameterized rock-paper-scissors game, we derive a condition under which communities of a given strategy form with either fixed or drifting boundaries. We show the effect of temperature increase on the underlying system, and identify surprising properties of this model through numerical simulations. 
\end{abstract}

\maketitle

\section{Introduction}\label{sec:Intro}

Traditional Darwinian evolution asserts that natural selection is driven by competition and mutation repeated over deep time, producing species mutually adapted to their environments and each other. Evolutionary game theory represents this process via (e.g.) the replicator or replicator-mutator equation, considering instantaneous reward from interaction as the measure of fitness. This approach, however, neglects the ability of organisms to store energy thereby insulating themselves from negative interactions. That is, fitness is (in some sense) measured by overall historical success as well as instantaneous success. In this paper, we study a spatial evolutionary game mechanism that incorporates mutation via a Boltzmann distribution but is mediated by stored (gross) winnings, or wealth. We derive conditions where communities form with high probability and study the resulting community structures theoretically and empirically in one dimension.

Evolutionary games have been studied extensively since Taylor and Jonker's original work \cite{TJ78}. In particular, the replicator equation and its variants have been studied extensively \cite{Schuster1983,H84,Friedman1991,HS98,W97,HS03,AMo04,cressman2014replicator,FS16,PG16,EGB16,GJW20}. The replicator equation does not admit natural mutation. Consequently, there has been investigation of the replicator-mutator variant \cite{K04,PL11,TS15,AC17}. While both the replicator and replicator-mutator equations operate in continuous time, there have also been several related studies of evolutionary dynamics in discrete time \cite{AMJ04,VRS11}; see the thorough (but dated) survey by Sigmund  \cite{sigmund1986survey}. Spatial evolutionary games have also been extensively studied \cite{ref:vickers1989,ref:vickers1991,ref:nowak,ref:cressmanVickers,ref:kerr,ref:andrew,ref:KT21,ref:roca,ref:RPSnetworks,griffin2021finite} with special focus on spatial rock-paper-scissors games, many of which do not use the replicator dynamic \cite{SMR14,SMR13,SMJS14,RMF08,RMF07,PR19,PR17,M10,HMT10}.

In this paper, we model both evolution and mutation processes in the rock-paper-scissors game by taking a Boltzmann distribution approach, in which players are most likely to imitate local competing strategies associated with higher accumulated winnings (wealth), and the temperature defines random fluctuations relative to this scale. As a result, players who accumulate wealth are less likely to change, effectively decreasing the importance of temperature locally in their region. We recently showed that an imitation dynamic based on wealth could lead to fairness in the Ultimatum Game \cite{CBG2021}.
Our model here is most related to the work in \cite{ref:nowak,nowak2004emergence} with follow-on analysis in \cite{traulsen2005coevolutionary,traulsen2006stochastic}. 
In the rock-paper-scissors (RPS) game, we observe in numerical simulations the formation of spatial {\it communities}, in which all players use the same strategy.
These results are consistent with those in \cite{ref:nowak}, however they occur for very different reasons. We derive payoff matrix conditions under which a clear phase transition occurs leading to the formation of communities, and provide conditions where communities cannot form. We also discuss scenarios in which drifting communities emerge. These are distinct communities that slowly migrate spatially as their boundaries continuously change. These results are shown to depend on the payoff matrix structure and the Boltzmann distribution governing strategy evolution.

The remainder of this paper is organized as follows: In \cref{sec:Model} we discuss the model formally. Theoretical results are presented in \cref{sec:Theory}. In \cref{sec:Numerics} we provide numerical examples of the possible behaviors identified in \cref{sec:Theory}. Conclusions are presented in \cref{sec:Conclusion}. 

\section{Model}\label{sec:Model}
Let $G=(V,E)$ be a graph with vertex set $V$ and edge set $E$. For a given vertex (player) $x \in V$, we denote 
by $\mathcal{N}[x]$ the graph neighborhood of $x$ \textit{including} $x$, i.e., the set of vertices adjacent to $x$ in $G$ and $x$ itself. We also denote by $\mathcal{N}(x)$ the graph neighborhood of $x$, excluding $x$ - the difference being square vs round brackets. 
For a symmetric game with $m$ strategies,
each vertex at time $t$ is described by its strategy index $\sigma(x,t) \in \{1,2,\dots,m\} = S$ and the sum of the total payoffs, the bank value $B(x,t)$. For convenience, we will use the indicator function:
\begin{displaymath}
\mathcal{I}(s,k) = \begin{cases} 1 &\text{if $k = s$} \\ 0 & \text{otherwise}.\end{cases}
\end{displaymath}
which we use to define a neighborhood bank value for each strategy $s \in S$
\begin{displaymath}
N_s(x,t) = \sum_{u \in \mathcal{N}[x]} \mathcal{I}(s,\sigma(u,t)) B(u,t).
\end{displaymath}
This local sum of bank values in the neighborhood of player $x$ will be used to evaluate the preference for each strategy $s$ in our model. 

Let $\pi:S \times S \to \mathbb{R}$ be the non-negative symmetric payoff function for the two-player game under study. Starting from randomly chosen initial strategies $\sigma(x,0)$ and 
$B(x,0) \equiv 0$,
the evolution of $\sigma(x,t)$ and $B(x,t)$ are defined for $t>0$ by the following:
\begin{enumerate}
\item All players play their neighbors, and update their bank values as:
\begin{displaymath}
B(x,t) = B(x,t-1) + \sum_{u \in \mathcal{N}(x)} \pi[\sigma(x,t-1),\sigma(u,t-1)].
\end{displaymath}
\item All players update their strategies probabilistically according to a (modified) Boltzmann distribution with constant $\beta$:
\begin{equation}
\Pr\left[\sigma(x,t) = s\right] = 
\frac{e^{\beta N_s(x,t)}}{\sum_{s\in S}e^{\beta N_s(x,t)}}.
\label{eqn:Boltzmann}
\end{equation}
\end{enumerate}
Analogously with statistical mechanics in physics, we define 
\begin{equation}
\beta = \frac{1}{kT},
\label{eqn:Beta}
\end{equation}
where $T$ is an ambient temperature quantifying the fluctuation scale, and $k$ would be the Boltzmann constant (we use $k=1$ throughout). 
% MORE HERE
The absence of a negative sign in the distribution does not indicate a negative temperature, but rather corresponds to the fact that, while in a thermodynamic system the lower energy states are more likely, choices in a game theoretic context will naturally favor the higher payoff (or higher wealth) states.
In our model, the neighborhood bank values $N_s(x,t)$ influence the time evolution of the probability distribution for each player's strategy choice $s$. This formulation is consistent with the bounded rationality formulation of game theory \cite{MW02}.

While the analysis presented in this paper can be similarly applied to general non-negative payoff functions, we focus here on the symmetric, positive payoff generalized rock-paper-scissors (RPS) game, with payoff matrix defined as:
\begin{equation}
 A(\delta,\epsilon) = \begin{bmatrix}
 1+ \delta & 0 & 2 + \epsilon \\
 2 + \epsilon & 1 + \delta & 0 \\
 0 & 2 + \epsilon & 1+ \delta\\
 \end{bmatrix}
 \label{eqn:Matrix}
\end{equation}
where $\epsilon \ge 0$ is the ``winning bonus", which allows us to study the role of larger payoffs for playing against a losing strategy, and $\delta \ge 0$ is the ``tie bonus", which we use to increase the reward for playing against the same strategy. With the additional condition $\delta < 1+ \epsilon$, the matrix retains the relative payoff inequalities of a generalized RPS game;
when $\epsilon = \delta = 0$, $A$ becomes the standard RPS matrix given in Weibull \cite{W97}. 
%We select $\epsilon$ and $\delta$ to ensure the payoff function remains non-negative since it is clear from the probability update rule that negative payoffs may prevent strategy fixation; i.e., the tendency of a player to repeatedly play the same strategy over and over again with increasing probability. 

\begin{figure*}[htbp]
\centering
\includegraphics[width=0.4\textwidth]{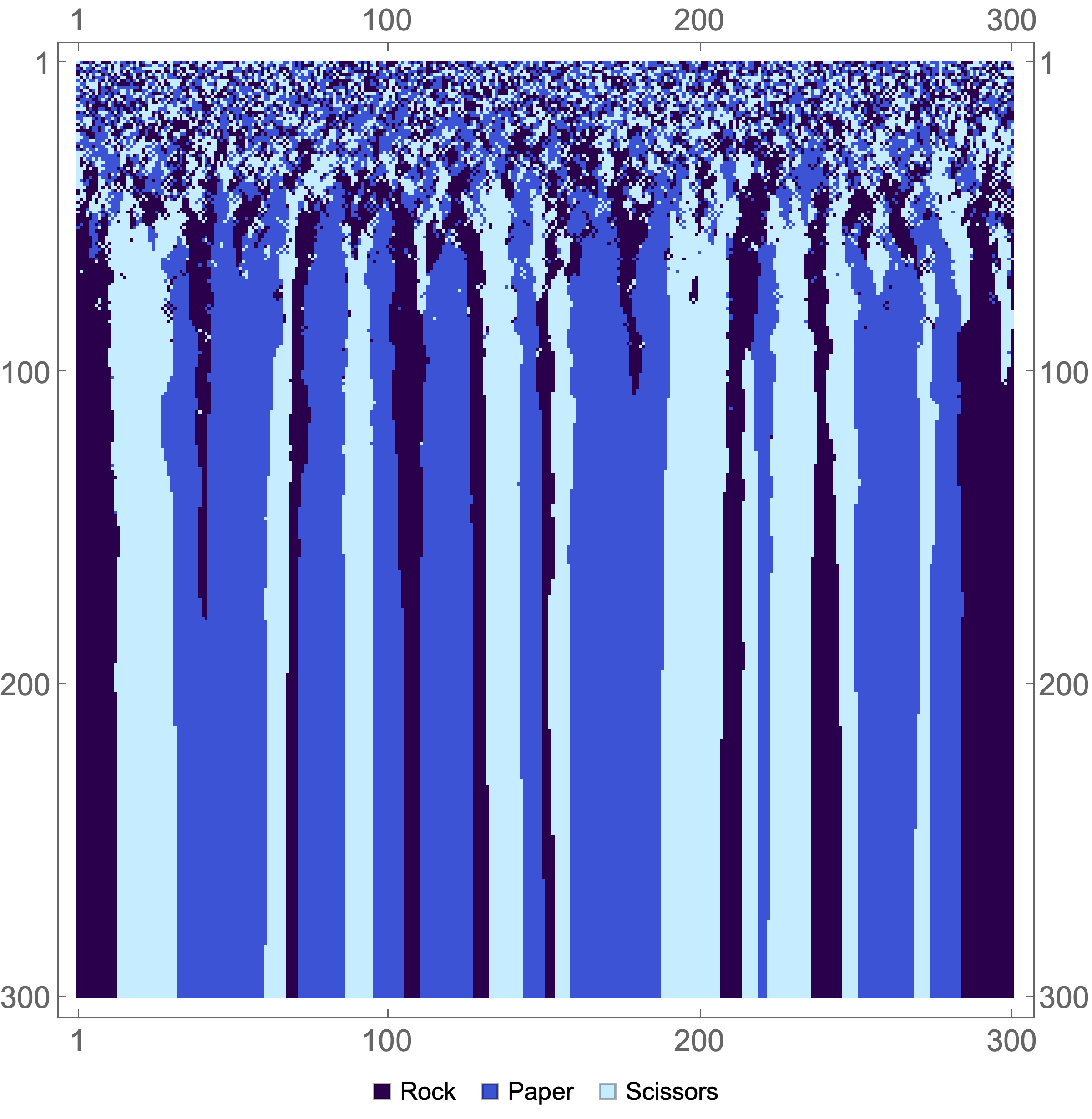}
\includegraphics[width=0.4\textwidth]{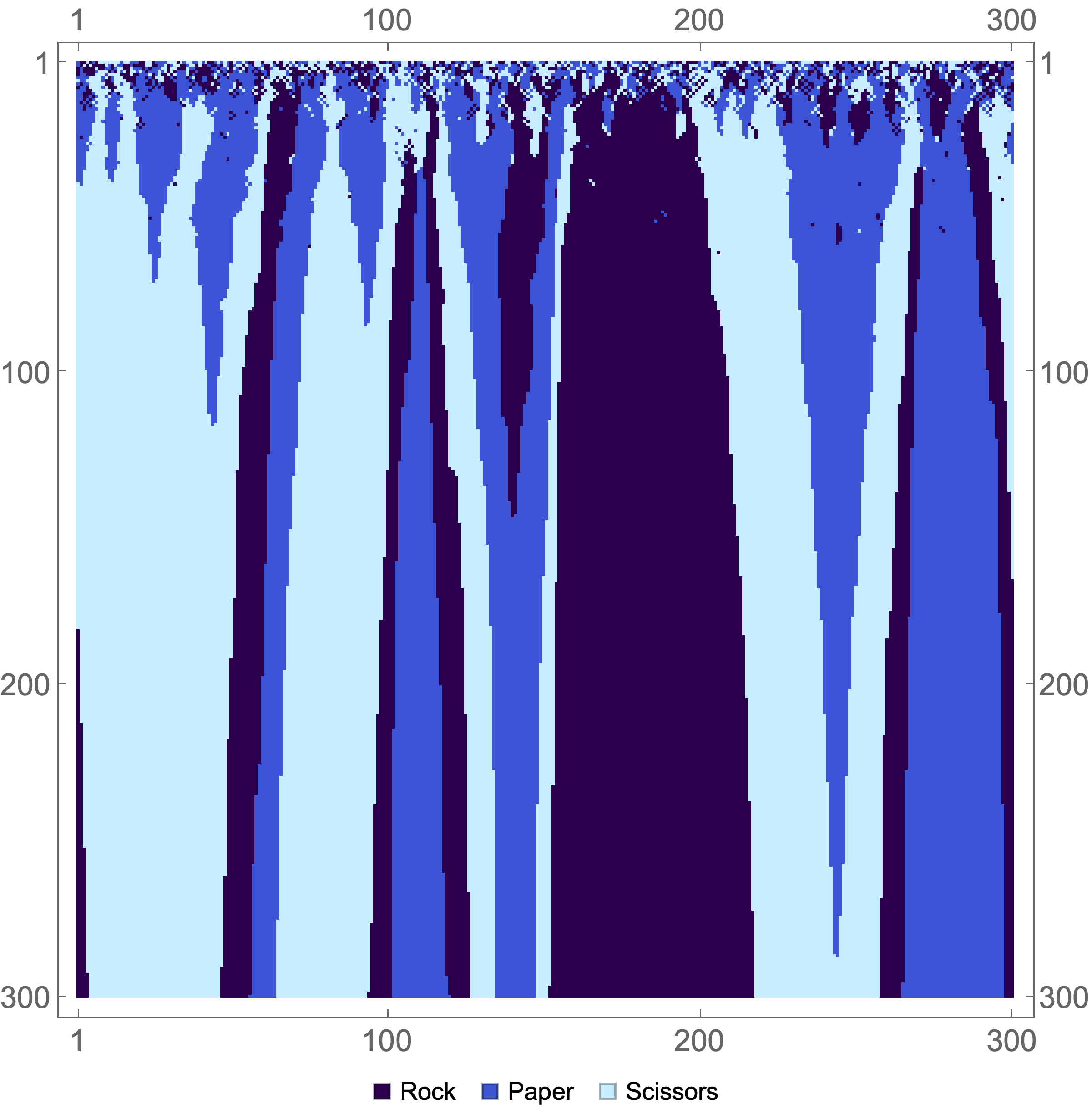}
\caption{Examples of community formation in our model: (Left) communities that become stationary $(\delta = 0.5,\epsilon = 0)$, (Right) communities that exhibit boundary drift $(\delta = 0,\epsilon = 16)$. }
\label{fig:Example}
\end{figure*}

For the remainder of this paper, we will focus on lattice graphs 
% with and 
without boundaries in one dimension; this is a cycle, and as such every player has two neighbors. %In the 1D case, the existence of a boundary does not have a major effect, but on a 2D lattice, the existence of the boundary has a major effect on the evolution of the strategy function $\sigma(x,t)$. 
An example of the dynamic behavior in our model is shown in \cref{fig:Example} for two different classes of $(\delta,\epsilon)$ values. 
Strikingly, we see that single strategy communities have formed along the lattice. In the left image, these communities are stationary while in the image on the right, they exhibit drift. We discuss community formation in detail in the following sections and provide numerical evidence showing the impact the relative values of $\epsilon$ and $\delta$ have on these communities.

\section{Theoretical Analysis}\label{sec:Theory}
 We now derive the probability that a given vertex $v$ on a cycle graph (a 1D lattice without boundary) will have the same strategy at time $t$ and $t+1$. We define a community as a group of two or more adjacent vertices that (i) share a strategy and (ii) have high probability of retaining that strategy over long time periods. For every community there are boundary members; i.e., those vertices that are adjacent to a vertex with a different strategy. We study how these boundary members evolve as a by-product of our analysis. We note that this dynamical system does not reach a strict equilibrium as there is always a probability of switching to any strategy, but if the probability to play the same strategy tends to 1 for each player, we can infer that the game tends toward stability over time.  
 
We now focus on the long-run probability of strategy fixation for the dynamical system we have described in \cref{sec:Model}. To do a thorough analysis on this game, we will study perturbations of the parameters in the payoff matrix given in \cref{eqn:Matrix}. 

In general, we assume $\delta$ and $\epsilon$ are small, but this isn't necessarily required. Note that when $\epsilon = 2 \delta$, we have a scaling of the original RPS form. We can break down the probabilities into cases based on what the neighbors of a focal vertex play with respect to that vertex; since the graph is a cycle, all vertices have exactly two neighbors. To focus on the probabilities of continuing to play the same strategy, extended for $n$ rounds into the future, we fix an initial time $t_0$ and define $N_s(t_0)$ as the bank value corresponding to the focal player's strategy.   

First, when both neighbors play the same strategy as the focal player (tie), the general form of the probability of repeating the same strategy next round is
\begin{equation*}
P_R = \frac{e^{4n(1+\delta)+n\phi}e^{N_s(t_0)}}{e^{4n(1+\delta)+n\phi}e^{N_s(t_0)}+2}
\end{equation*}
where $\phi$ is specified by how the neighbors perform against their {\it other} neighbor. There are six different possibilities
\[
\phi =
\begin{cases}
0 & \text{both neighbors lose}\\
2(2+\epsilon) & \text{both neighbors win}\\
2(1+\delta) & \text{both neighbors tie}\\
(1+ \delta) & \text{one ties, one loses}\\
(2+\epsilon) & \text{one wins, one loses}\\
(3+ \delta + \epsilon) & \text{one wins, one ties}\\
\end{cases}
\]
Now that we have a complete expression of this case, we can give a description of what it represents. If $n=0$, the expression gives the probability of playing the same strategy in next round, given $N_s(t_0)$. When $n$ is nonzero, the value is the probability of playing the same strategy in the $n+1^\text{th}$ round given $N_s(t_0)$ and assuming that locally no player changes strategy over the first $n$ rounds.  

We can use this expression to study how this probability behaves as $n \to \infty$. If it tends to 1, we can conclude it becomes increasingly likely that the focal player continues to play this strategy, but there is always a nonzero probability that the player will choose a different strategy. If it tends to 0, we can say with probability 1 that the player will eventually choose a different strategy. As noted previously, this is only one case. The other situations are described below.  

The previous case described a setting when both neighbors played the same strategy as the focal player. When that isn't the case, there are many different variations which can be grouped into five cases around how the focal player performs: (a) lose on both sides, (b) win on both sides, (c) win on one and lose on the other, (d) win and tie, and (e) lose and tie. The case of a tie on both sides was the case just considered. Let $P_R$ denote the probability that the focal player repeats. Then we have:

\paragraph{Case 1:} The focal individual loses on both sides. The general form of the probability to play the same strategy is
\[
P_R = 
\frac{e^{-2n(2+\epsilon) +n \phi}e^{\Delta N(t_0)}}{e^{-2n(2+\epsilon) +n \phi}e^{\Delta N(t_0)} + 1 + e^{-N_c(t_0)}e^{-2n(2+\epsilon) + n \phi}}
\]
where again $\phi$ is the contribution of the secondary neighbors. Here, and for the following cases, we will define 
$$
\Delta N(t_0) = N_s(t_0) - N_c(t_0),
$$
where $N_c(t_0)$ is the total bank value corresponding to the (complementary) strategies that the focal player isn't playing. Similar to the previous case, $\phi$ is then
\begin{equation}
\phi =
\begin{cases}
0 & \text{both neighbors lose}\\
-2(2+\epsilon) & \text{both neighbors win}\\
-2(1+\delta) & \text{both neighbors tie}\\
-(1+ \delta) & \text{one ties, one loses}\\
-(2+\epsilon) & \text{one wins, one loses}\\
-(3+ \delta + \epsilon) & \text{one wins, one ties}\\
\end{cases}
\label{eqn:nu1}
\end{equation}
These cover all potential situations in the lose on both sides case.

\paragraph{Case 2:} Win on both sides. This case is fortunately very similar to the previous one, except now the focal individual gets a payoff on each play. Thus the general form becomes
\[
P_R = 
\frac{e^{2n(2+\epsilon) +n \phi}e^{\Delta N(t_0)}}{e^{2n(2+\epsilon) +n \phi}e^{\Delta N(t_0)} + 1 + e^{-N_c(t_0)}e^{n \phi}}
\]
with the form of $\phi$ given by \cref{eqn:nu1}. 

\paragraph{Case 3:} Win on one side, lose on the other. This is the most complicated of the different cases as each neighbor is playing a different strategy, so we cannot write the general form simply. Instead, we will leave it in the general form, and denote $N_W(t_0)$ as the bank of the winning neighbor and  $N_L(t_0)$ of the losing. We then have the general form as
\[
P_R = 
\frac{e^{n(2+\epsilon)}e^{N_s(t_0)}}{e^{n (2+\epsilon)}e^{N_s(t_0)} + e^{N_W(t_0)}e^{n(2+\epsilon)+n \phi_W} + e^{N_L(t_0)}e^{n \phi_L}}
\]
In this particular case, we denote $\phi_W$ and $\phi_L$ separately, so we can express them simply as 
\[
\phi_W, \phi_L = 
\begin{cases}
0 & \text{lose}\\
2+\epsilon & \text{win}\\
1+\delta & \text{tie}\\
\end{cases}
\]
In special cases this equation does simplify.

\paragraph{Case 4:} Win on one side, tie on the other.  The general form is given by
\[
P_R = 
\frac{e^{n(4+2\delta+\epsilon) + n\phi}e^{\Delta N(t_0)}}{e^{n(4+2\delta+\epsilon)
+n \phi}e^{\Delta N(t_0)} + 1 + e^{-N_c(t_0)}e^{-n \mu}}
\]
This case is more delicate then the previous cases because anything won by the neighbor playing the same strategy contributes positively to the probability, while anything won by the other neighbor contributes negatively. Thus we must have cases for each potential combination and can no longer appeal to symmetry. For utmost clarity, a \textit{teammate} will refer to the neighbor playing the same strategy as the focal player, and we will specify what is happening to this teammate. In the previous three cases, all $\phi$ counted against the focal player's strategy. Since one of the neighbors now plays the same strategy, this must be accounted for. The defined $\phi$ will still treat both neighbors, but we now introduce $\mu$ to handle only the portion which is contributed by the other, non-teammate strategy. The different $\phi$ are
\begin{equation}
\phi = 
\begin{cases}
0 & \text{both lose,win, or tie}\\
1+\epsilon - \delta & \text{teammate wins, other ties}\\
\delta - 1 - \epsilon & \text{teammate ties, other wins}\\
2+ \epsilon & \text{teammate wins, other loses}\\
-2 - \epsilon & \text{teammate loses, other wins}\\
1+ \delta & \text{teammate ties, other loses}\\
-1-\delta & \text{teammate loses, other ties} \\
\end{cases}
\label{eqn:nu2}
\end{equation}
with $\mu$ being
\begin{equation}
\mu = 
\begin{cases}
0 & \text{lose}\\
1+ \delta & \text{tie}\\
2+ \epsilon & \text{win}\\
\end{cases}
\label{eqn:mu}
\end{equation}
Note that $\mu$ can be discerned from $\phi$ and the other information, but notationally this is simpler.

\paragraph{Case 5:} Lose on one side, tie on the other. This case is very similar to the previous, with a slightly different general form given by
\[
P_R = 
\frac{e^{n(2\delta-\epsilon) + n\phi}e^{\Delta N(t_0)}}{e^{n(2\delta-\epsilon)
+ n \phi}e^{\Delta N(t_0)} + 1 + e^{-N_c(t_0)}e^{-n(2+\epsilon)-n \mu}}
\]
Here $\phi$ and $\mu$ are still given by \cref{eqn:nu2,eqn:mu}, the same as in the previous case.

We have given a complete characterization of every potential case that can arise on a cycle (1D lattice with no boundary), which also allows for the study of limit probabilities under the assumption that nothing changes. The logic used to arrive at these formulae can easily be extended to the 1D lattice with boundary by considering the cases where the focal individual has only one neighbor. In what follows, we see how the asymptotic convergence manifests, and which situations will never be stable in the limit.
 
\subsection*{Asymptotic Behavior}
Studying the system numerically suggests that as bank accounts increase, communities begin to form (see \cref{sec:Numerics} below). For the sake of the analysis, we will assume that communities have already formed, and will focus on players at the boundary between two communities. As we are studying the RPS game, one of these players must be winning against the other. We focus on the losing player, as their bank value grows the slowest out of all players around them. If the two communities are large enough, then the probability for this losing player to play the same strategy next round is given by
 \begin{equation}
     P_R = \frac{e^{n(2\delta -\epsilon)}e^{\Delta N(t_0)}}{e^{n(2\delta -\epsilon)}e^{\Delta N(t_0)}+ 1 + e^{-N_c(t_0)}e^{-n(3+\epsilon+\delta)}}
 \end{equation}
 We focus on this case because it is the ``worst case'', and thus will reveal how stability depends on the relationship between $\epsilon$ and $\delta$. 

 First, suppose $2\delta > \epsilon$. Then, if nothing locally changes each round, we can study the limit of this probability, which is
 \begin{equation}
 \lim_{n \to \infty} \frac{e^{n(2\delta -\epsilon)}e^{\Delta N(t_0)}}{e^{n(2\delta -\epsilon)}e^{\Delta N(t_0)}+ 1 + e^{-N_c(t_0)}e^{-n(3+\epsilon+\delta)}} = 1
 \label{eqn:StationaryLimit}
 \end{equation}
 There is some difficulty with this result, as any change during play breaks it, and there is always a positive probability for such a change to occur. What we do see is that the probability converges exponentially to 1, so even though we don't have convergence in finite time, in practice the probability to change quickly approaches extremely small values.

Considering the alternate case, that is $\epsilon>2\delta$, then the limit goes in the other direction. We have 
 \begin{equation}
  \lim_{n \to \infty} \frac{e^{n(2\delta -\epsilon)}e^{\Delta N(t_0)}}{e^{n(2\delta -\epsilon)}e^{\Delta N(t_0)}+ 1 + e^{-N_c(t_0)}e^{-n(3+\epsilon+\delta)}} = 0
  \label{eqn:TransientLimit}
 \end{equation}
 The conclusion we can derive from this is much more precise than in the previous case. Since the probability tends to 0 exponentially, we know for certain that the focal player will eventually change their strategy as the product of these probabilities converges even more quickly to 0. How this manifests in practice is through transient community boundaries. The bank value for the neighbor's strategy eventually surpasses the bank value for the focal player's strategy, and thus the probability tends toward 1 that the focal player will adopt this strategy. This happens at every boundary, and drives the gradual shifting of communities in the domain, as seen for example in Fig.~\ref{fig:Example}R.

The remaining case, where $2\delta = \epsilon$, is the most interesting. This case represents the classic RPS game. 
% In particular the standard RPS we study here falls within this case. 
Here limit is
 \begin{equation}
 \lim_{n \to \infty} \frac{e^{\Delta N(t_0)}}{e^{\Delta N(t_0)}+ 1 + e^{-N_c(t_0)}e^{-n(3+\epsilon+\delta)}} = \frac{e^{\Delta N(t_0)}}{e^{\Delta N(t_0)}+1}
 \label{eqn:RPSLimit}
 \end{equation}
This limit will tend towards a fixed probability which is neither 0 nor 1. So again we conclude that the focal player will eventually change strategy, as the product of this probability for each round goes to zero.  
 
 However, what is most interesting is the time dependence of the limiting probability $P_R$ for this case. Since we can assume that $N_c(t_0)$ is fairly large, then:
 \begin{equation*}
 e^{-N_c(t_0)}e^{-n(3+\epsilon+\delta)} \approx 0,
 \end{equation*}
 so the probability to play the same strategy is essentially constant, with a value dependent solely on $\Delta N(t_0)$. This manifests itself in a surprising way. Unlike the $2\delta > \epsilon$ case, here $P_R < 1$, so there is no tendency toward fixation. And unlike the $\epsilon>2\delta$ case, the boundary probabilities don't tend toward 0 either, so there is no guarantee that the boundaries will drift. Instead, we find a combination of the two. Note that $\Delta N(t_0)$ is simply the difference in bank values at the beginning time $t_0$ of our analysis. Since the boundary is guaranteed to move at some point (as $P_R$ doesn't converge to 1), there will be a time $t_1$ when the boundary will move. If $\Delta N(t_1)>\Delta N(t_0)$ (where note these $\Delta N$ are evaluated for different players, since the boundary has changed), then the probability to play the same strategy has increased for the losing player at the boundary. Therefore this probability can in fact tend toward 1, as in the first case, but it only does so by moving the boundary itself, which is the phenomenon of the second case. Thus the asymptotics of the $\epsilon = 2\delta$ case is a boundary case which dynamically combines the other two scenarios.

\section{Numerical Results}\label{sec:Numerics}

We now study the three conditions discussed in the previous sections numerically. In all simulations, we fix $T = 100$ ($\beta = 0.01$), which affects the rate at which communities form.
% *** REF TO FOLLOW UP PAPER??

%The first set of simulations were run for 400 rounds. In practice, the constant $k$ only has an effect on how quickly the bank values reach a value large enough to facilitate community formation. In this setup, communities begin to form when players have around 130 bank value. 

%\subsection{Illustrative Examples}
%%Insert new figures and discussion here.
\subsection{Probability of Stationarity}
\label{sec:Prob} 
\begin{figure}[htbp]
\centering
\includegraphics[width=0.45\columnwidth]{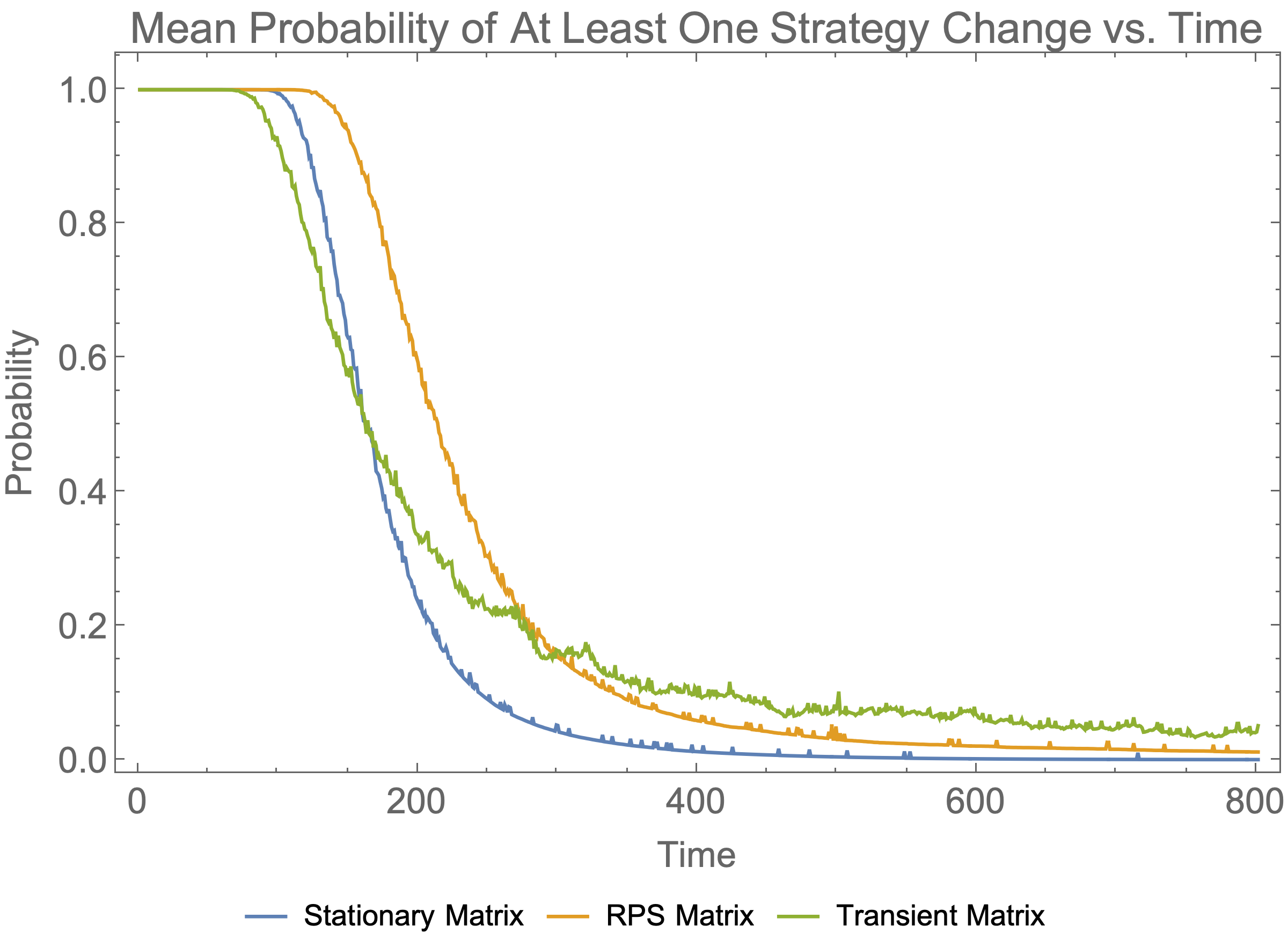}\quad
\includegraphics[width=0.45\columnwidth]{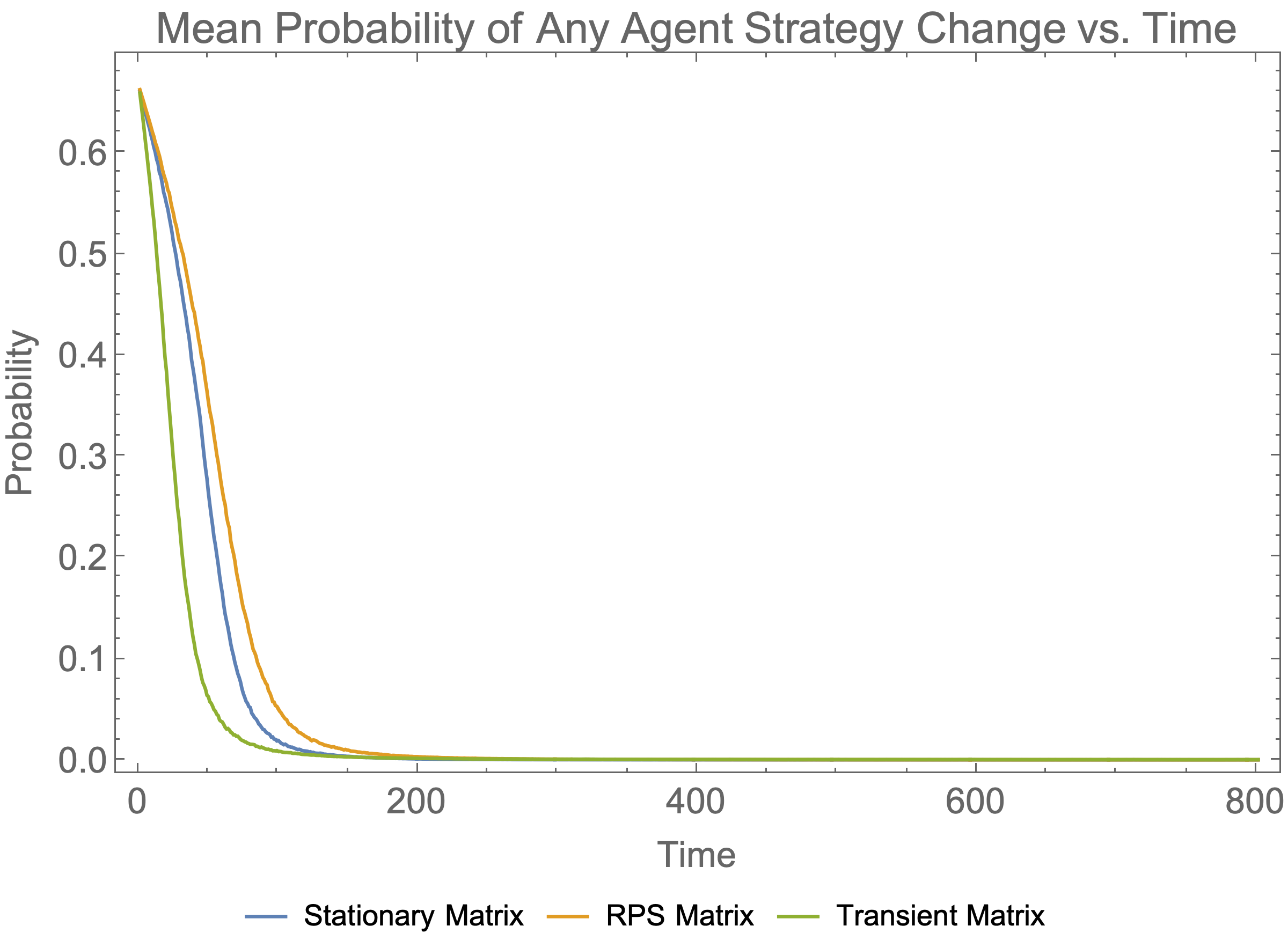}
\caption{(Left) The mean probability of at least one player changing over time decreases the most in games with the stationary matrix as expected. (Right) The mean probability of any player changing over time decreases rapidly to zero.}
\label{fig:MeanPN}
\end{figure}

Consider the $p_R(v,t)$, the probability that player (vertex) $v$ maintains its strategy from round $t$ to round $t+1$. To study this function we used a cycle (domain) of size 300. Payoff matrices were defined as follows:
\begin{enumerate}
\item For the stationary matrix ($\epsilon < 2\delta$) we use parameters $\delta = 0.5$ and $\epsilon = 0$. 
\item For the standard RPS matrix ($\epsilon = 2\delta$) we use parameters $\delta = \epsilon = 0$.
\item For the transient matrix ($\epsilon > 2\delta$) we use parameters $\epsilon = 2$ and $\delta = 0$. 
\end{enumerate}
We ran 100 realizations of each scenario to measure $p_R(v,t)$, and used this to compute the probability that at least one agent would change strategy from time $t$ to $t+1$ as:
\begin{equation*}
p_1(v,t) = 1 - \prod_v p_R(v,t).
\end{equation*}
The simulation was stopped at $T_\mathrm{max} = 800$. The numerically determined mean $\langle{p_1}\rangle$ is shown in \cref{fig:MeanPN}.

\begin{figure*}[ht!]
\centering
\includegraphics[width=0.95\textwidth]{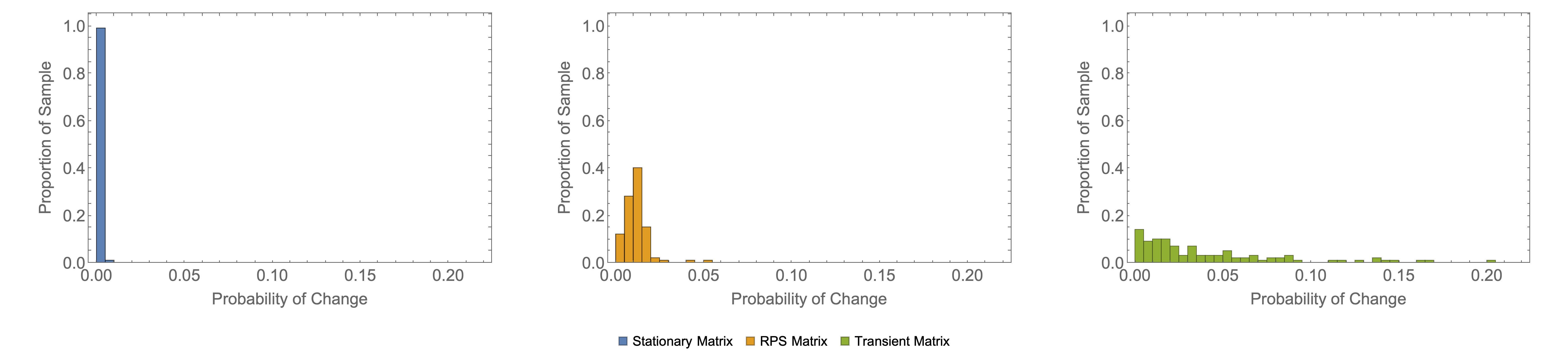}
\caption{Distributions of the probability of at least one player changing from round $800$ to round $801$. (Left) Stationary matrix, (Middle) standard RPS matrix, (Right) Transient matrix.}
\label{fig:PN800Histogram}
\end{figure*}

We observe the expected decrease in time in the probability that at least one lattice point will change strategy with $\langle{p_1}\rangle \to 0$ for the stationary matrix and $\langle{p_1}\rangle > 0$ for the transient matrix at $T_\mathrm{max}$. This is consistent with the theoretical analysis showing community drift should occur in the transient case. The RPS matrix exhibits behavior between transient and stationary cases. Histograms for $p_1(v,T_\mathrm{max})$ over all sample runs are shown in \cref{fig:PN800Histogram}. 
Again we see the greatest dispersion in the transient matrix case as the probability of maintaining strategy should go to zero - see \cref{eqn:TransientLimit}.

Locality is the key aspect of this spatial system. An individual player sees the strategies and bank values of their direct neighbors, and is only effected by their neighbors and their neighbors' neighbors. This locality leads to an interesting result. Even if one player has a bank value much larger than all the others, that player's strategy will only necessarily be dominant in a small community centered around that player. This results from the lack of redistribution of bank value, so the dominant player will always be dominant in their local area, but that local area does not extend far. %This affects the probability of strategy change.

Interestingly, while \cref{fig:MeanPN} shows the expected differences between the three matrix conditions, the mean probability that any one individual changes strategy is close to zero as time goes to infinity. This can be explained because most of the players are within a community and so have a very large probability to play the same strategy. The boundary players probabilities are not extreme enough to influence this average from the dominant majority. What these curves do show is the community formation rate is largely independent of the limiting behavior in these three cases, although in the transient case communities do form faster as the winning bonus $\epsilon=2$ is large enough to influence this rate. The communities are then subject to drift while they are stationary when $\epsilon < 2\delta$. Surprisingly, it appears that communities form slowest in the case when $\epsilon = \delta = 0$, which is a property of the parameters themselves. Communities solidify as the bank values increase, and the rate of this increase depends on how large $\epsilon$ and $\delta$, and thus the entries of the payoff matrix, are. Therefore the rate of community formation is a function of the magnitudes of the entries of the payoff matrix, while the long time community behavior is a function of the relationship between these payoffs.

\subsection{Community Sizes and Scaling}
To determine how the domain size affects both the number of communities and the community sizes, we ran simulations with domain (cycle) sizes ranging from 150 - 1200 vertices. We ran the simulations 1000 times for each community size for each matrix type. Each simulation was terminated after 600 rounds, and the number and size of each community were determined. 
We define a community to be a set of at least two adjacent vertices in the domain that all have the same strategy. Single vertices not belonging to a community did not occur in our sample after $\sim$600 rounds. 
%Here a community is a set of vertices all sharing a strategy that are adjacent to each other in a graph.
%Communities of size 1 are not considered communities and did not occur in our sample after 600 rounds.
\begin{figure*}[!ht]
\includegraphics[width=0.3\textwidth]{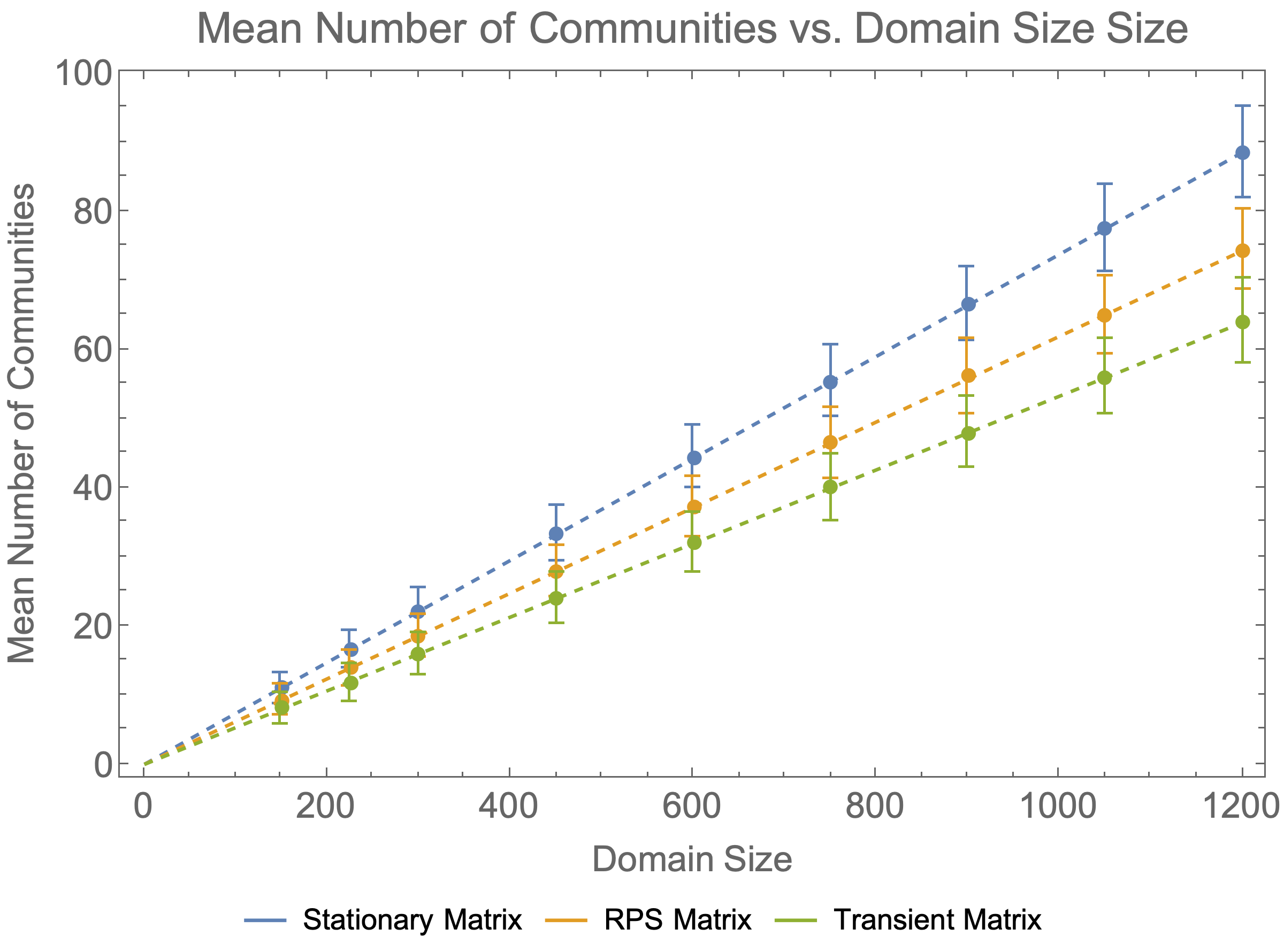}
\includegraphics[width=0.285\textwidth]{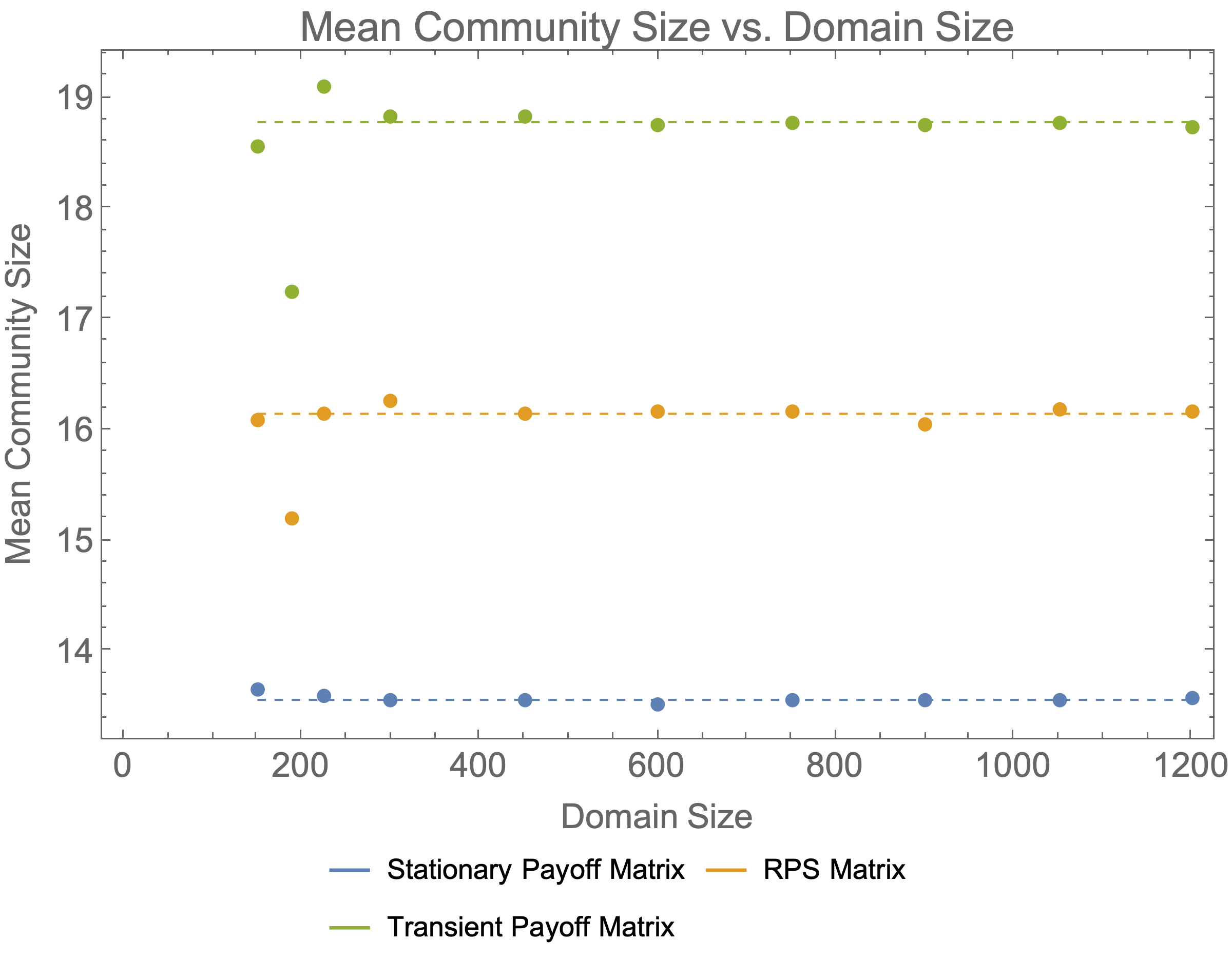}
\includegraphics[width=0.304\textwidth]{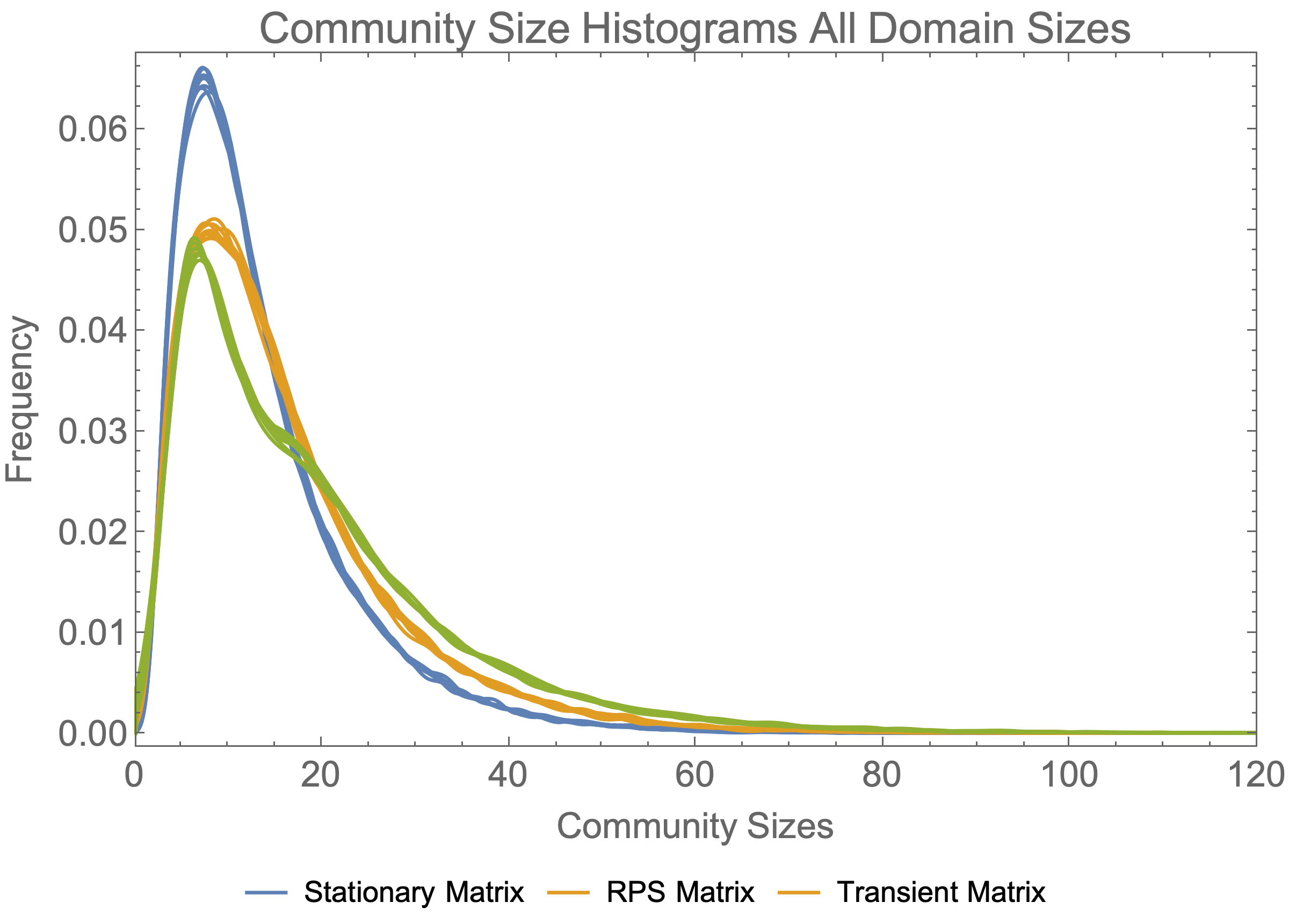}
\caption{(Left) Average number of communities relative to the domain size. Error bars represent one standard deviation. (Middle) Average community size relative to the domain size. (Right) Combined histograms of all community sizes. Note that the histograms for simulations with the same matrix type but different domain sizes are nearly identical.}
\label{fig:Histograms}
\end{figure*}
\cref{fig:Histograms} (Left) and (Middle) show the relationships of domain size with average number of communities and mean community size, respectively; \cref{fig:Histograms} (Left) shows a linear relationship between the average number of communities and the domain size. As a result of this linear growth, we expect to see a constant mean community size over all domain sizes, as seen in \cref{fig:Histograms} (Middle). \cref{fig:Histograms} (Right) shows smoothed histograms for all simulations. Note that the histograms for simulations with the same matrix type are nearly identical but differ among the matrix classes. 

We note in \cref{fig:Histograms} (Left) that the stationary case has a larger average number of communities than either standard rock-paper-scissors or the transient case. This is further illustrated in \cref{fig:Histograms} (Right) which shows the community sizes are smaller (distribution is more to the left) with a thinner tail when compared to the other cases. We hypothesize that this is because the communities in the stationary case solidify earlier, when there are many small communities from the initial few rounds when the probability to play any strategy is essentially uniform. On the other hand, the transient case has a smaller number of average communities. This is consistent with our speculation from the previous section, where over time communities are eliminated, but often none are brought into existence, so in general we would expect that there would be fewer larger communities. 

We tested this empirically using a small (200 replication) simulation with $(\epsilon,\delta) = (18,0)$ and $(\epsilon,\delta) = (0,9)$. While this is outside the region we consider for parameters, we note the results are consistent with our previous results. \cref{fig:OtherEpsDelta} shows the confirmation of the hypothesis that, as $\epsilon$ increases with fixed $\delta$, the number of communities per domain size increases, while as $\delta$ increases with fixed $\epsilon$, the number of communities decreases. 
\begin{figure}[!ht]
\centering\setcounter{figure}{4}
\includegraphics[width=0.5\columnwidth]{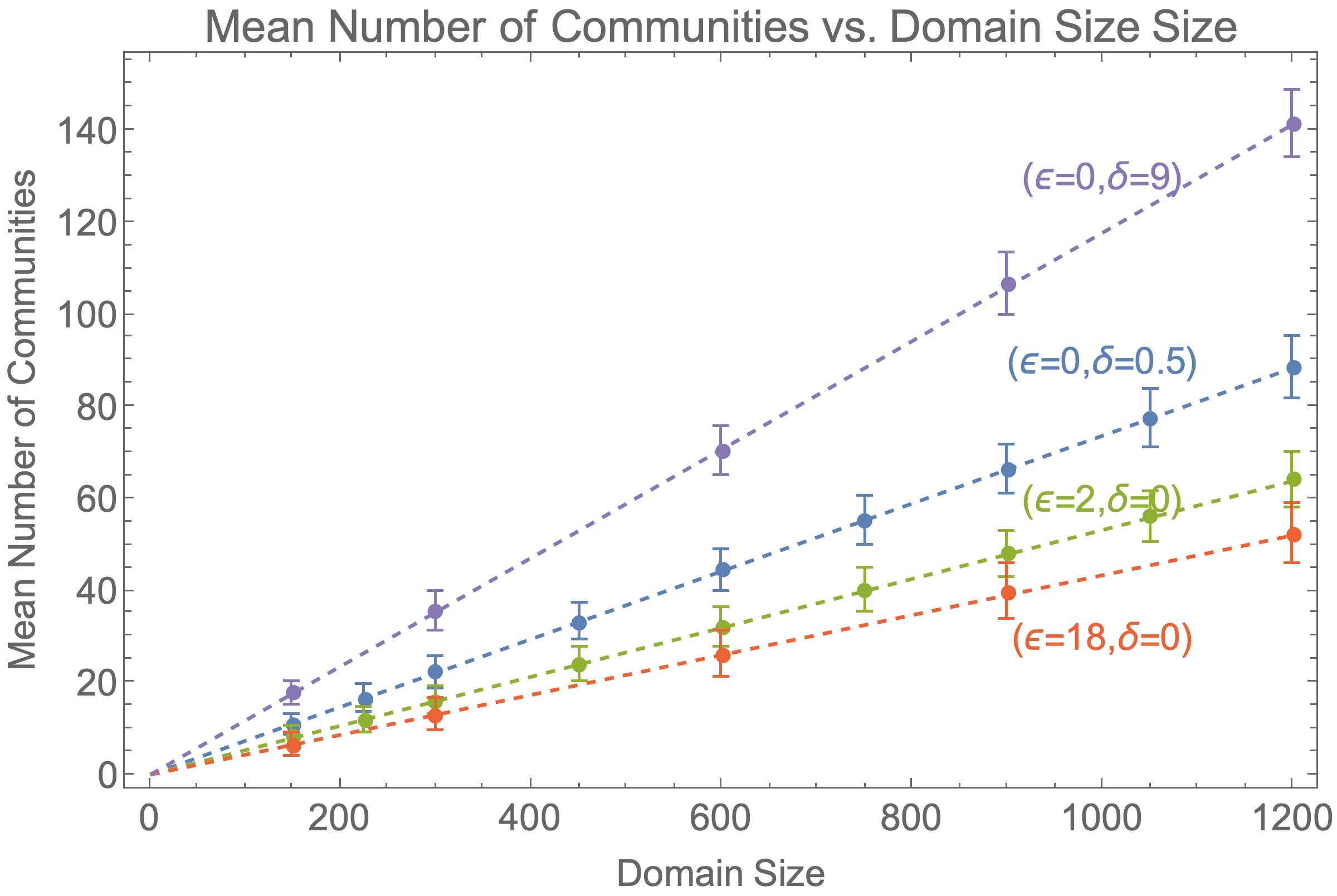}
\caption{We include two additional values for $\epsilon$ and $\delta$ in the stationary and transient cases illustrating their effect on community formation.}
\label{fig:OtherEpsDelta}
\end{figure}
Because each $(\epsilon,\delta)$ combination seems to lead to a natural mean community size $\lambda$ (which is the inverse of the slope shown in \cref{fig:OtherEpsDelta}), we know that:
\begin{align*}
\lim_{\epsilon \to \infty} \lambda(\epsilon,\delta) \geq 2\\
\lim_{\delta \to \infty}\lambda(\epsilon,\delta) \leq \Omega,
\end{align*}
where $\Omega$ is the domain size. If $\lambda(\epsilon,\delta) = \Omega$, then there is a single monoculture community. In future work we will attempt to determine the structure of $\lambda(\epsilon,\delta)$ and to understand why a natural community emerges at all, since it is not immediately clear that this should be so from the model structure.

%\cref{fig:Histograms}(b) shows that the average community size is independent of the environment size, which is unsurprising given the linear relationship in \cref{fig:Histograms}(a), as knowing the average community size and the environment size gives the average number of communities. Mathematically, 1/mean community size is the slope of the line in \cref{fig:Histograms}(a).

%
%Figure \cref{fig:Histograms}(c) shows the distributions of community size from all simulations for each matrix type. Studying more cases of $\epsilon$ and $\delta$ is required determine how these depend on the precise matrix details, especially the interesting shape of the transient case distribution. 
%
%Figure \cref{fig:Histograms} (d) and (e) look at the distributions of number of communities in different ways from \cref{fig:Histograms}(a) and (c). \cref{fig:Histograms}(d) focuses on an environment size of 1200 for the three matrix types. These distributions are centered where we expect given the lines in \cref{fig:Histograms}(a), but it is interesting that they are roughly symmetric about the average. It isn't intuitive why large communities are just as likely as small communities relative to the mean. \cref{fig:Histograms}(e) shows the distributions for the stationary case matrix \ref{eqn:convmat} for three community sizes. 
%
%Future work will focus on studying how these types of distributions change relative to environment size, especially for small numbers of cells. 

\subsection{Effects of A Linear Temperature Increase}
Because all payoffs in the general RPS games considered here are non-negative, there is an overall trend for the vertex bank values to increase linearly with time. Thus the importance of any fixed $kT$, which defines the scale of fluctuations in the Boltzmann distribution, diminished with time, which may lead to a ``freezing in" of communities. In order to compensate for this, and study systematically the interplay of random fluctuations with spatial community structure, we impose a linear temperature ramp, of the form
\begin{equation*}
T = T_0 + at,
\end{equation*}
where $T_0 = 100$, consistent with the previous sections. This allows us to adjust the value of $a$ so that the thermal fluctuations can keep pace with the growth of $N_s$. As expected, we observe a kind of ``melting" of the communities to different extents, dependent on $a$, as illustrated in \cref{fig:Melting}.
\begin{figure*}[!htbp]%%This is just for figure spacing.
\centering\setcounter{figure}{5}%%Hard rest of counters to make this ordered correctly.
\includegraphics[width=0.3\textwidth]{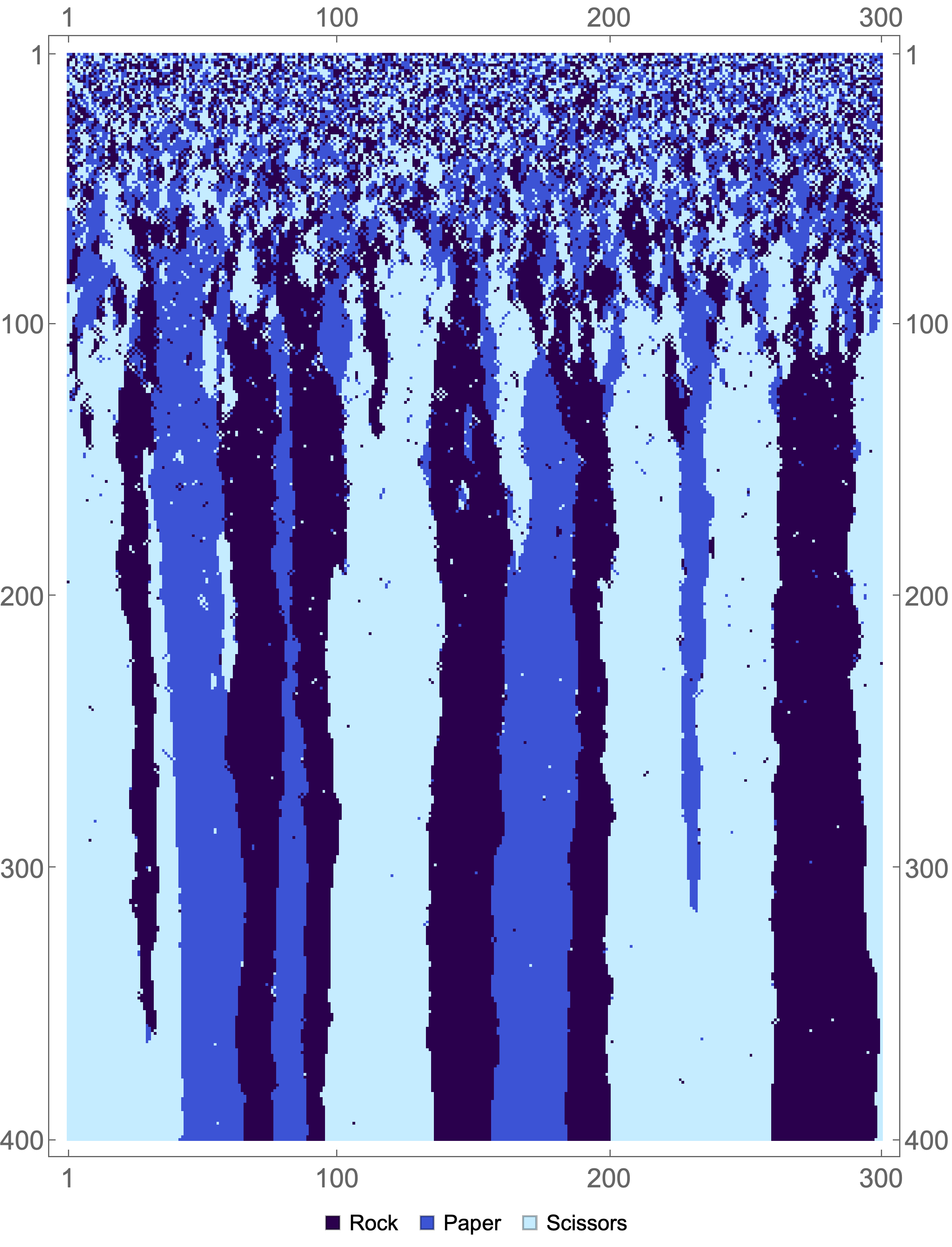}\quad
\includegraphics[width=0.3\textwidth]{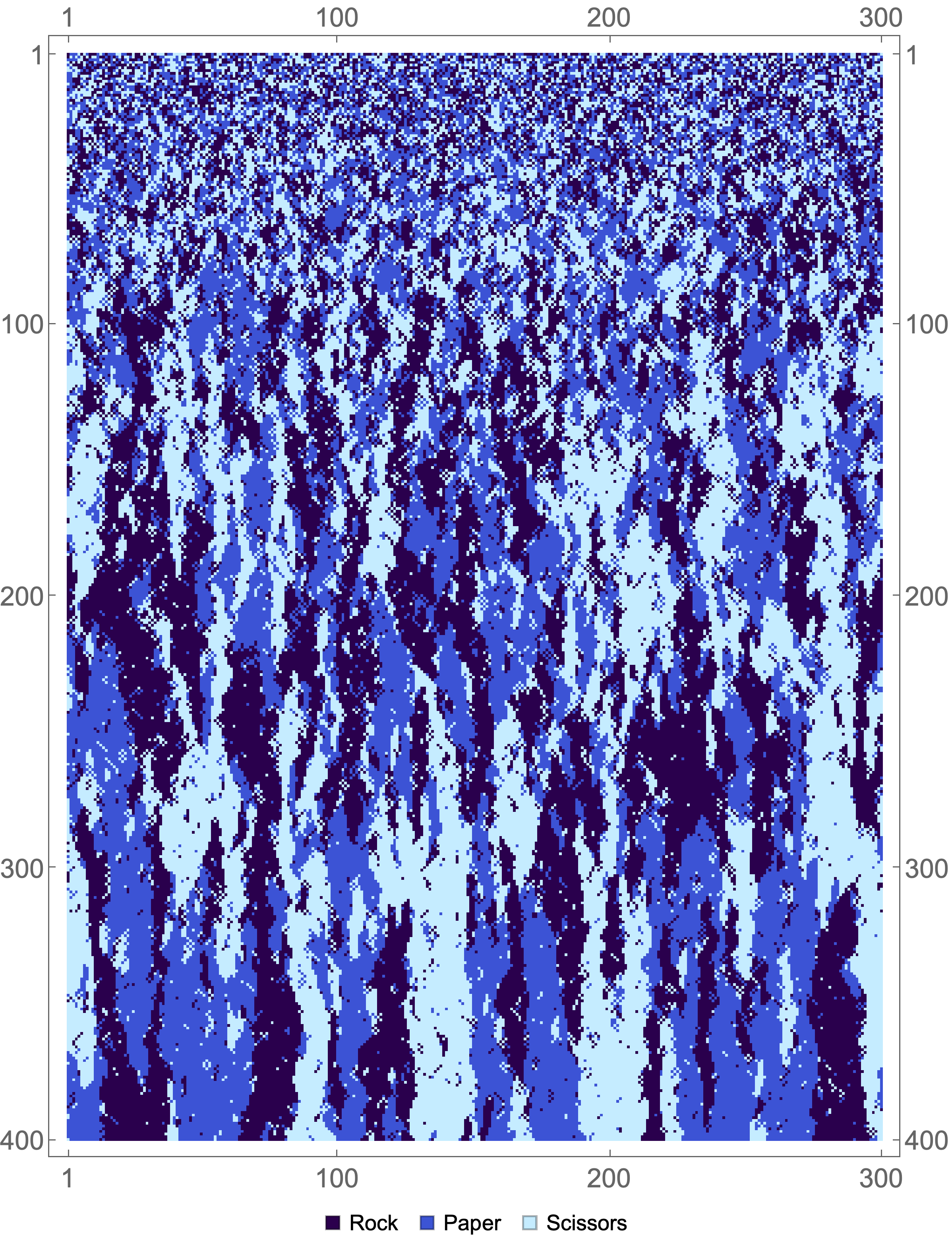}\quad
\includegraphics[width=0.3\textwidth]{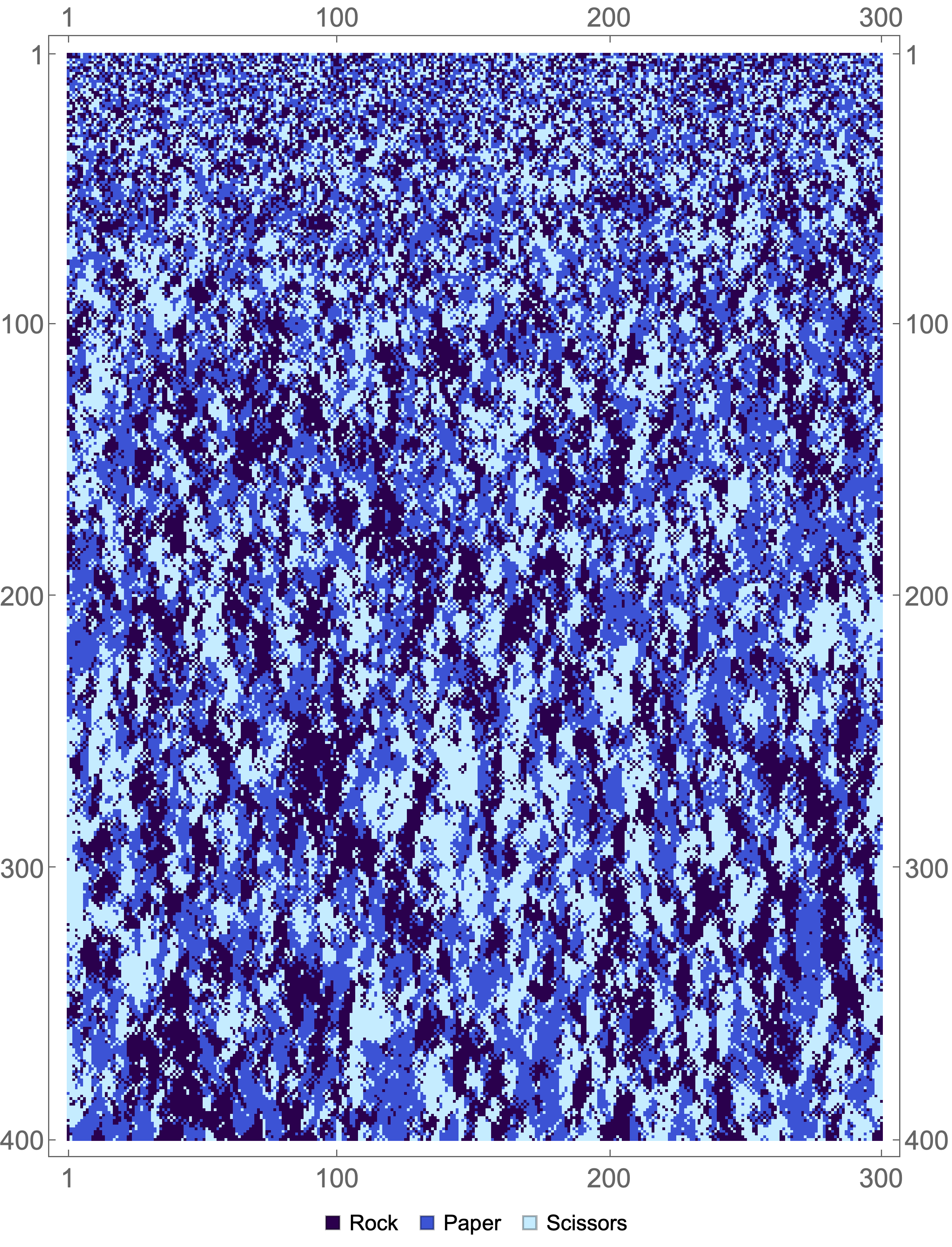}\quad
\caption{The effect of a linear temperature ramp on the formation of communities: (Left) $a = 0.8$, (Middle) $a = 1.5$, (Right) $a = 2$. As $a$ is increased, the more rapidly increasing temperature causes a melting, in which communities become smaller and more stochastic in dynamics.}
\label{fig:Melting}
\end{figure*}
As $a$ is increased from zero, the effect of thermal fluctuations becomes more pronounced, and communities seem to become smaller. For large values of $a$, no communities are observed, as the overall increase of bank values cannot outpace the rate of temperature increase. We study this in the $2\delta > \epsilon$ case, for which stable communities form at $a = 0$. We ran 100 replications with varying values of $a$ using a fixed % community - typo?
domain size of 300, with the matrix parameters as given in Section \ref{sec:Prob}.

\begin{figure}[htpb]
\centering
\includegraphics[width=0.45\columnwidth]{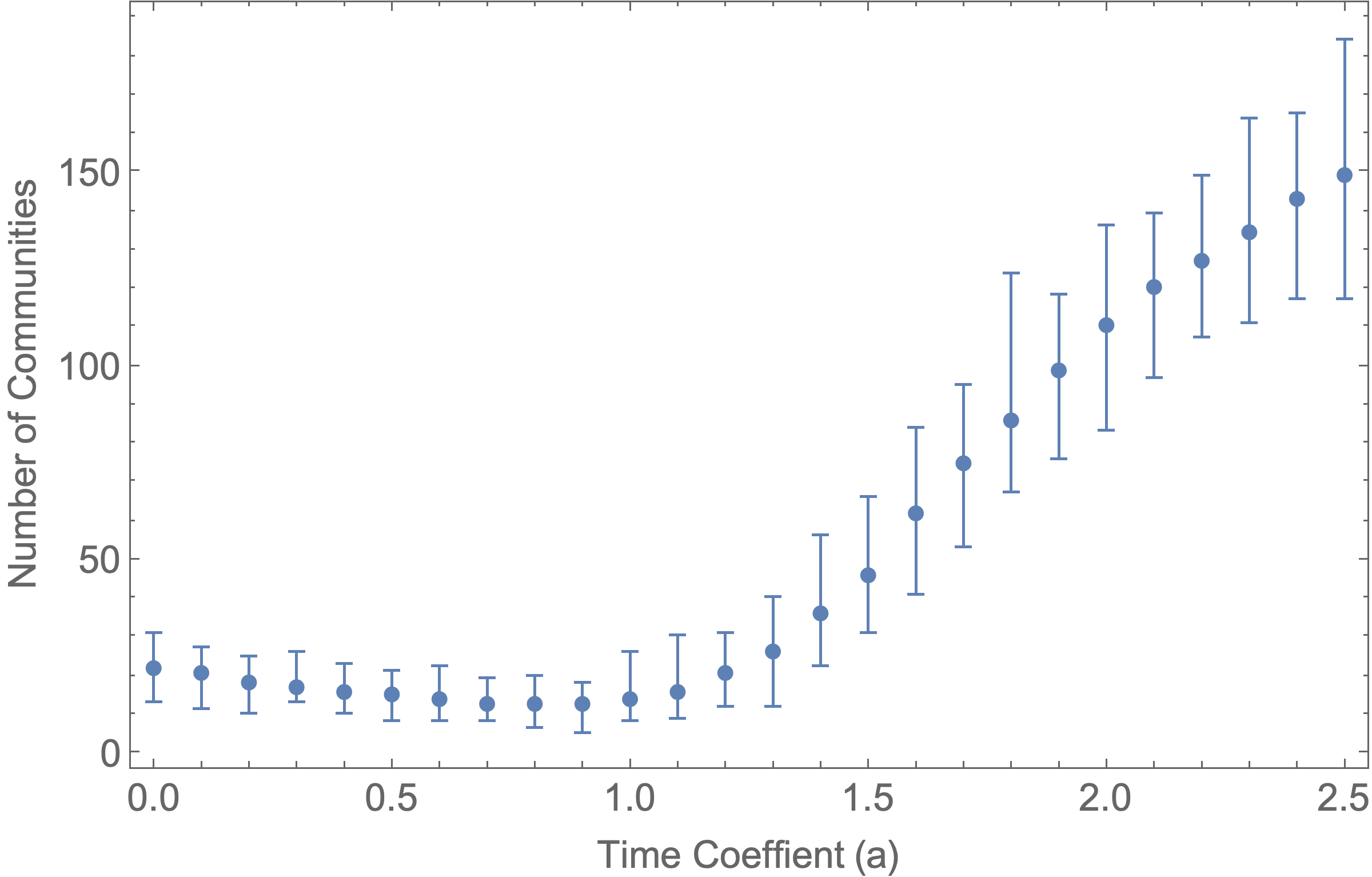} \quad
\includegraphics[width=0.45\columnwidth]{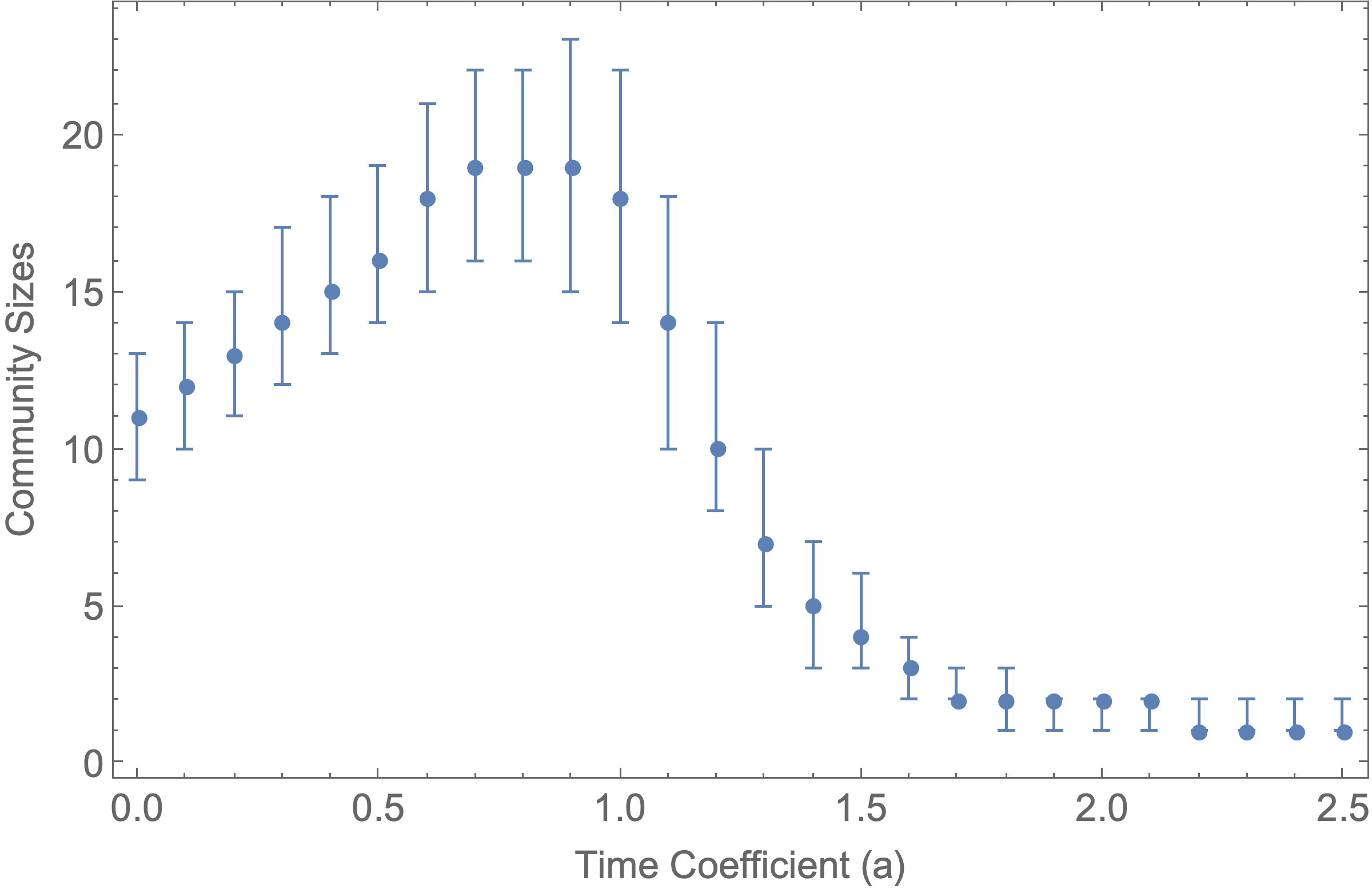}
\caption{For increasing temperature ramping $a$, the median number of communities (Left) and median community size (Right) initially decrease (respectively increase), as smaller communities are annealed and merge into larger ones. As $a$ is further increased, the community structure disappears (melting). Error bars: the full range of the distribution (Top), and 40\%-60\% of the distribution (Bottom).}
\label{fig:MeltingEffect}
\end{figure}

Results are shown in \cref{fig:MeltingEffect}, in terms of the median number of communities and median size, with error bars representing the range of the distribution (see caption). We see that that median community count reaches a minimum at $a \approx 0.8$, with a corresponding maximum in the median community size at the same value of $a$. Using a Mann-Whitney test we can see the size of the communities when $a = 0$ and when $a = 0.8$ is statistically different at well beyond $7\sigma$. We hypothesize that the continued presence of fluctuations due to the temperature ramp allow smaller communities, which would have otherwise survived, to merge with other larger communities. This ``annealing" effect results in an increased median community size, and decreased number of communities. 
It also suggests the possibility that a linear temperature ramp followed by a fixed temperature could provide a means to control community sizes.

At larger $a$ values, the relatively larger fluctuations overcomes the stabilization of larger communities, as the median community size decreases while the number of communities increases; eventually, the community structure melts entirely (\cref{fig:Melting}R).

 \section{Conclusion}\label{sec:Conclusion}
 In this paper we have studied a dynamical system arising from evolution in a repeated game on a lattice, where choice of strategy is mediated by a Boltzmann distribution. We compute the probability of strategy fixation within this dynamical system and use it to explain the formation of communities in three classes of the rock-paper-scissors matrix. We show that two of these cases have community boundaries becoming (effectively) fixed as time goes to infinity, thus leading to stable communities. In the third case we show that the communities are (slowly) transiting across the lattice domain. 
 
%%Rewrite?
Studying the distributions of the number and size of the communities formed in this model has revealed some surprising relationships. Future work will focus on explaining why the average size of the community is independent of the domain size and only depends on the choice of matrix parameters. Additionally, characterizing how this dependence arises from the dynamics is of significant interest. Our results may offer new interpretations and possibilities for modeling species diversity and coexistence in biological systems such as lichen communities \cite{mathiesen2011}. 
We have also shown that increasing temperature can cause a decrease in the number of communities (leading to larger communities) for slow temperature increases. For larger temperature increases, a larger number of (highly transient) smaller communities form as a result of temperature effects. Understanding this relationship would be intriguing. In particular, it would be intriguing to determine whether the system is a glass and if so what its more specific properties are. Determining other system characteristics and community properties, especially in higher dimensions, is also of interest. From the perspective of the statistical physics of social systems \cite{castellano2009}, studying wealth redistribution schemes in this context may also provide additional insights.

%Finally, in other future work, it would be interesting to see whether other thermodynamic properties can be leveraged to understand this behavior.   %An interesting change to explore would be to allow some form of resource redistribution within the communities after each round. This way a single dominant player could spread their influence beyond their local area. 
 
 \section*{Acknowledgements}
 The authors were supported in part by the National Science Foundation Grant DMS-1814876.
 
C. Olson is also supported by the National Science Foundation Graduate Research Fellowship Program under Grant No. DGE1255832. Any opinions, findings, and conclusions or recommendations expressed in this material are those of the author(s) and do not necessarily reflect the views of the National Science Foundation.

\bibliography{ThermoGame-Chaos}
\end{document}